\documentclass[%
 reprint,
 amsmath,amssymb,
 aps,
 prl,
 longbibliography,
 lengthcheck,%
]{revtex4-1}

\usepackage{graphicx}% Include figure files
\usepackage{dcolumn}% Align table columns on decimal point
\usepackage{bm}% bold math
\usepackage{hyperref}% add hypertext capabilities
\begin{document}

\preprint{APS/123-QED}

\title{First principles derivation of NLS equation for BEC with cubic and
quintic nonlinearities at non zero temperature. Dispersion of linear waves.}% Force line breaks with \\

\author{P. A. Andreev}
\email{andreevpa@physics.msu.ru}
 \affiliation{Department of General Physics, Physics Faculty, Moscow State
University, Moscow, Russian Federation.} %Lines break automatically or can be forced with \\

\begin{abstract}
In this work we presented a derivation of the quantum hydrodynamic
equations for neutral bosons. We considered short range
interaction between particles. This interaction consist binary
interaction $U(\textbf{r}_{i},\textbf{r}_{j})$ and three particle
interaction $U(\textbf{r}_{i},\textbf{r}_{j},\textbf{r}_{k})$,
the last one does not include binary interaction between particles. From
the quantum hydrodynamic (QHD) equations for Bose-Einstein
condensate we derive nonlinear Schr\"{o}dinger equation. This
equation includes the nonlinearities of third and fifth degree. It is at zero temperature.
Explicit form of the constant of three-particle interaction was taken.
First of all, developed method we used for studying of dispersion of linear waves. Dispersion
characteristics of linear waves were compared for the cases. It were of
two-particle interaction in approximation third order to
interaction radius (TOIR) and three-particle interaction, at zero temperature. We consider
influence of temperature on dispersion of elementary excitations.
For this aim we derive a system of QHD equations at non-zero
temperature. Obtained system of equation is an analog of
well-known two-fluid hydrodynamics. Moreover, it is generalization
of two-fluid hydrodynamics equations due to three-particle
interaction. Evident expressions of the velocities
of the first and second sound via the concentrations of superfluid
and noncondesate components is calculated.

\end{abstract}

\maketitle

\section{\label{sec:level1}I. Introduction}

At the theoretical investigation of Bose-Einstein condensation
(BEC) with account of the three-particle interaction (TPI)
the nonlinear Schr\"{o}dinger
equation (NLSE) is used. This equation contain  cubic and quintic nonlinearities
~\cite{Kovalev FNT 76,Barashenkov PLA 88}. For describing inelastic scattering imaginary part of
interaction constants is used
in  ~\cite{Abdullaev PRA 01}. Three-particle interaction leads to
forming of dimers. This process leads to loss of atoms from
BEC ~\cite{Bedaque PRL 00,Jack PRL 02}.

 Opportunity of obtaining of the Gross-Pitaevkii (GP) equation from the microscopic many
particle Schr\"{o}dinger equation is demonstrated in ~\cite{Erdos
PRL 07}. The GP equation is the example of NLSE. In ~\cite{Andreev PRA08} the direct derivation of the GP
equation from many particle Schr\"{o}dinger equation has done.

In article ~\cite{Andreev PRA08} authors derived NLSE for system of
bosons being in the state of BEC, with the short-range interaction
potential, in third  order in interaction radius. If one keeps
only the term of the first order in interaction radius then one
gets well-known GP equation from many-particle Schr\"{o}dinger
equation. There taking into account only potential of binary
interaction. At derivation authors do not use suggestion about density
of considered system. Therefore, there occur interaction whether
by means of scattering or during finite interval of time.
Presented in ~\cite{Andreev PRA08} derivation does not exclude
interaction of a few particles at the same moment.

In this article we use the same method for studying
of both a three particle interaction (TPI) and influence of
temperature on dynamic of neutral Bose particles.

Fundamental and detailed distribution of influence of non-zero
temperature on BEC dynamic is presented in paper ~\cite{temp BEC},
where were considered the two-particle interaction only. The
influence of the noncondensate atoms on BEC dynamic is considered
in ~\cite{temp BEC}, but dynamic of noncondesed atoms do not included.
Interference of superfluid and normal
components usually described by two-fluid hydrodynamic
~\cite{Landau 6}. Present day examples of using of two-fluid
hydrodynamics can be found in papers ~\cite{Griffin arXiv 2011,Griffin 98 4044,Griffin 97,Griffin 98 4695}.
Connection of two-fluid hydrodynamics with the
kinetic equations is presented in papers ~\cite{Griffin 98 4695,Griffin 04}.
In cited papers authors used Boltzmann  like
kinetic equation with the collision term.

Due to studying of three particle interaction the ground
state energy density in second order in $\sqrt{\rho a^{3}}$ is
calculated in ~\cite{Braaten EPJ B 99,Lee PR 57 v105,Lee PR 57
v106,Yang Physica 60,Wu PR 59,Hugenholtz PR 59,Sawada PR 59},
where $\rho$ is the particles density, $a$ is the s-wave
scattering length. The first calculation of the constant under the
logarithm is presented in ~\cite{Braaten EPJ B 99}. The constant
depends on $a$ and from the parameter
described the lower energy $3\rightarrow 3$ scattering of the
particles.

 The explicit three-body contact potential
for a dilute condensed Bose gas is derived from microscopic theory
in ~\cite{Kohler PRL 02}. The derivation is based on the quantum
expectation values of products of single mode annihilation and
creation operators. The three-body coupling constant exhibits the
general form predicted by Wu ~\cite{Wu PR 59}. It depends on
s-wave scattering length $a$, defined via binary potential like in
~\cite{Wu PR 59}. For describing the properties of BEC at zero
temperature, with three-body interaction, in ~\cite{Kohler PRL 02}
there were used nonlinear Schr\"{o}dinger equation contained
nonlinearity of fifth degree.

Here we present some physical effects there is in BEC due to TPI.
Stability properties of BEC with TPI are considered in ~\cite{Wamba PRE
08,Abdullaev PRA 01,Marklund EPJ.B. 05}. The modulation instability of the
BEC trapped in an external parabolic potential, is investigated in
~\cite{Wamba PRE 08}. The explicit time-dependent criterion
for the modulation instability of the condensate was established.

 The influence of initial conditions on
stability of BEC was studied in ~\cite{Abdullaev PRA 01} using a Gaussian
variational approach and numerical simulations, in
three-dimensional trapped BEC. Abdullaev et.al. ~\cite{Abdullaev
PRA 01} discussed the validity of the criterion of stability
suggested by Vakhitov and Kolokolov. In the works ~\cite{Wamba PRE
08,Abdullaev PRA 01} for studying of properties of BEC with TPI
there were used nonlinear equation with cubic and
quintic nonlinearities.

Abdullaev et.al. ~\cite{Abdullaev PRA 05} discussed localized
ground states of BEC in optical lattices with attractive and
repulsive TPI. For this aim a quintic
nonlinear Schr\"{o}dinger equation has used. In ~\cite{Abdullaev
PRA 05} the existence of unstable localized excitations which are
similar to Townes solitons of the cubic nonlinear Schr\"{o}dinger
equation in two dimensions is shown.

Marklund et. al. ~\cite{Marklund EPJ.B. 05} found the Vlasov-like
equation for BEC. They used Wigner function and NLSE with
cubic-quintic nonlinearities.

Under the condition, that two-particle scattering length tends to
zero, it may be realized by manipulating external magnetic field,
far off-Feshbach resonance ~\cite{Bloch RMP 08}, $\Upsilon$ tends
to zero too. In this situation main role has TOIR and TPI. In this
case TOIR and TPI are compared.

There are various generalizations of GP equation. Here we interesting
in nonlocal NLSE. We present brief review of known nonlocal NLSE generalization of GP equation.
For studying of BEC the NLSE with nonlinearities containing
spatial derivatives of wave functions is used. The NLSE for BEC
with nonlocal nonlinearity was obtained in ~\cite{Rosanov PL.A.
02}. In that article  NLSE contains second spatial derivatives of
square of wave function module.  In ~\cite{Andreev PRA08} NLSE in
third order on interaction radius (TOIR) approximation is derived.
In this case NLSE is integro-differential equation with spatial
derivatives of wave functions. Comparison between nonlocal
nonlinearity ~\cite{Rosanov PL.A. 02} and TOIR ~\cite{Andreev
PRA08} for two dimensional space is considered in ~\cite{Andreev
Izv.Vuzov. 10}. Local Lagrangian density, for two-body and
three-body effective short-range interaction, contained dependence
on $(\psi^{*}\psi)^{2}$, $(\psi^{*}\psi)^{3}$ and
$\nabla(\psi^{*}\psi)\nabla(\psi^{*}\psi)$ is presented in
~\cite{Braaten PRA 01}. There are given a determination of the
strength of the three-body contact interaction for various model
potentials.

Generalization of GP equation to the large-gas-parameter regime is
suggested in ~\cite{Fabrocini PRA 01}. It is NLSE with
nonlinearities of fourth degree.

In this article we use method of quantum hydrodynamic (QHD) of
many particles system developed in works ~\cite{Maksimov
99,Maksimov 01,Andreev PRA08}. In ~\cite{Maksimov 99}
 equation of QHD of charged particles system was found in
external electromagnetic field taking into account Coulomb
interaction of particles. In ~\cite{Maksimov 01} for system of
charged and spinning particles there were derived equations of
quantum hydrodynamic, i.e., continuity, balance of momentum,
magnetic moment and energy. In ~\cite{Maksimov 99} there were
developed methods of calculation of many-particle functions
arising in equations of QHD. In ~\cite{Andreev PRA08} there was
developed method of quantum hydrodynamics for system of bosons,
fermions and mixtures, with short-range interaction. In this
article there were given corresponding equations of continuity and
balance of momentum. On the base of system QHD equations the NLSE
describing the dynamics of boson-fermion mixtures is obtained.
Particular case NLSE for bosons noticed in ~\cite{Andreev PRA08}
is the GP equation.

For boson systems with two- and three-particle interaction, the
problem of finding the quantum stress tensor can be solved and
conditions of its existence can be clarified. Here, we will limit
ourselves by the development of the quantum stress tensor and show
that under the standard assumptions this tensor can be transformed
so that the momentum balance equations for bosons coincide with
the analogous NLSE contained cubic and quintic nonlinearities.
Using this momentum balance equation derivation, we can obtain the
equation (for wave function in medium or the order parameter),
which is NLSE with nonlinearities of fifth degrees. Quantum stress
tensor for boson systems is symmetric. It in first order in
interaction radius consists of two parts, namely, the term
coincides with interaction via binary interaction potential and
the term coincides with TPI which does not
contain interaction by means binary potential. The first term
leads to equation coinciding with the analogous GP equation. The
second term includes explicit expression for three particle
interaction constant and coincides with the term analogous to the
nonlinearity of fifth degrees in NLSE.

In dilute alkali gases, when interaction may be considered like
scattering process, the constant of interaction in first order on
interaction radius (FOIR) has the form
$\Upsilon=-4\pi\hbar^{2}a/m$; $a$- is scattering length. In TOIR
approximation there arise the second constant of interaction
$\Upsilon_{2}$. In general case, parameter $\Upsilon_{2}$ is
independent from $\Upsilon$ and has to considered like
supplementary. In ~\cite{Andreev PRA08} an approximate  estimate
of $\Upsilon_{2}$ via $\Upsilon$ is considered.

In the paper ~\cite{Andreev arxiv 11 1} were shown the account of
interaction up to TOIR leads to finding of new physical effects in BEC.
In ~\cite{Andreev arxiv 11 1} were found new type of solitons in BEC.
Frequency dependence of eigenwave in boson-fermion mixture in TOIR
is derived in ~\cite{Andreev PRA08}. Therefore, generalization of
Bogoliubov spectrum ~\cite{Bogoliubov N.1947},
~\cite{L.P.Pitaevskii RMP 99} was obtained in TOIR. Also,
analytical dependence $\omega(k)$ for degenerate fermions is
obtained. Using system of QHD equations (derived in ~\cite{Andreev
PRA08}) in works ~\cite{Andreev Izv.Vuzov. 09 1,Andreev Izv.Vuzov.
10} was investigated dynamics of nonlinear wave in TOIR. In
~\cite{Andreev Izv.Vuzov. 10}  analytical solution for bright
soliton is obtained in uniform BEC within the TOIR approximation.
In ~\cite{Andreev Izv.Vuzov. 09 1} nonlinear eigenfrequency shift
in uniform BEC is analytically investigated within the TOIR
approximation.

At experimental investigation of BEC  in magnetic traps work with
number of particles $N$ is order $10^{4}$-$10^{5}$. At
quantum-mechanical description of such system it is necessary to
solve Schr\"{o}dinger equation determining wave function that
depends from $3N$ coordinate and time. Wave process, process of
transfer, exchange by energy and momentum at interaction take
place in three-dimensional physical space. In this connection it
is necessary to convert Schr\"{o}dinger equation to equation is
determined dynamic of functions in three-dimensional physical
space. This task is solved with method of quantum hydrodynamics,
further development of this method is made in this work.

Moreover, in this work problem of finding of method
allowed to build NLSE on the base system of QHD equations is solved.

Contribution of noncondensate particles was shown to be arise
in QHD equations for BEC and degenerate fermions ~\cite{Andreev
PRA08}.

Our paper is organized as follows. In Sect.2 we  derive QHE's for
BEC from many-particle Schr\"{o}dinger equation with two- and
three-particle interaction. In Sect.3 we calculate quantum stress
tensor due to three-particle interaction. In Sect.4 we derive
system of QHD equations for system of bosons at non-zero
temperature. We consider separate dynamic of two type of bosons,
it is particles in BEC state and noncondensate bosons. We obtain
the continuity equations and Eiler equations for each type of
bosons. This equations include interaction between different type
of bosons. The two- and three-particle interaction are included
too. In Sect.5 from QHD equations finding in sect.3 we obtain NLSE
with nonlinearity of third and fifth degrees. We give special
attention for interaction BEC with temperature excited particles,
but part of this presented in the Appendix 3. In Sect.6 we
construct equations for BEC including two-particle interaction in
FOIR, TOIR and contribution of first term of three-particle
interaction. On this background we obtain frequency dependence of
elementary excitation on wave vector for BEC with two- and
three-particle interaction at zero temperature limit. In Sect.7 we
study the dispersion of elementary excitations at non-zero
temperature. We obtain the velocity of the first and second sounds
due to TPI. In Sect.8 brief summary of
obtained results are presented.

\section{\label{sec:level1}II. Derivation of quantum hydrodynamic equations}

In the case, when we can present interaction between three
particles $V_{3}(\textbf{r}_{i}, \textbf{r}_{j}, \textbf{r}_{k})$
 in the form
$$V_{3}(\textbf{r}_{i}, \textbf{r}_{j},
\textbf{r}_{k})=U(\textbf{r}_{ij})$$
 \begin{equation}\label{tp three part int} +U(\textbf{r}_{ik})+U(\textbf{r}_{jk})+U(\textbf{r}_{ij},\textbf{r}_{ik},\textbf{r}_{jk}),\end{equation}
where $\textbf{r}_{ij}=\textbf{r}_{i}-\textbf{r}_{j}$,
$U_{ij}=U(\mid \textbf{r}_{ij}\mid)$-is the binary interaction
potential,
 $U_{ijk}=U(\mid
\textbf{r}_{ij}\mid,\mid \textbf{r}_{ik}\mid,\mid
\textbf{r}_{kj}\mid)$-is the three particle interaction potential
that does not contain combination of binary potentials $U_{ij}$.
Collisions of three particles in dilute gases is rare. But in the
case of liquid helium there is interaction of more than two
particles at the same moment. Moreover, this interaction has not
form of scattering.

 Let us
consider the system of N Bose-particles with a short-range
potential. The Hamiltonian of the system under consideration has
the form:
$$\hat{H}=\sum_{i}\frac{1}{2m_{i}}\hat{p}^{\alpha}_{i}\hat{p}^{\alpha}_{i}+\sum_{i}V_{ext}(\textbf{r}_{i},t)$$
\begin{equation}\label{tp ugam ht}+\frac{1}{2}\sum_{i,j\neq i}U_{ij}+\frac{1}{6}\sum_{i,j\neq i;k\neq i,j}U_{ijk} ,\end{equation}
 where
$\hat{p}^{\alpha}_{i}=-\imath\hbar\nabla_{i}$-is
the momentum operator of the i-th particle, $m_{i}$-is the
mass of the i-th particle. Let us consider the three-particle
interaction nonequivalent to combination of binary interaction
consider by $U_{ij}$. Interaction of three or more particles at the same time by means of binary potential is described by $U_{ij}$ and have no connection with TPI.

Atoms in the traps are kept by means interaction of their magnetic
moments with trapping magnetic field. In that theory this
interaction implicitly take into account by means $V_{i, ext}$. In
the magnetic field of third particle may exist effect like
Feshbach resonance ~\cite{Bloch RMP 08}. Due to this effect
scattering length of two-particles interaction is changed. It may
be considered like mechanism of three particle interaction, which
is not provided by binary potentials, only (\ref{tp three part
int}).

The concentration of particles in the vicinity of the point
$\textbf{r}$ of the physical space is determined as the operator
$\sum_{i=1}^{N}\delta(\textbf{r}-\textbf{r}_{i})$ averaged over
the quantum-mechanical states:
\begin{equation}\label{tp udefn ht}n(\textbf{r},t)=\int
dR\sum_{i}\delta(\textbf{r}-\textbf{r}_{i})\psi^{+}(R,t)\psi(R,t),\end{equation}
where $dR=\prod_{i=1}^{N}d\textbf{r}_{i}.$

Differentiating this function over time and using the
Schr\"{o}dinger equation with the Hamiltonian (\ref{tp ugam ht}),
we derive the continuity equation, in which the current density
vector appears in the form:
$$j^{\alpha}(\textbf{r},t)=\int dR\sum_{i}\delta(\textbf{r}-\textbf{r}_{i})\frac{1}{2m_{i}}$$
\begin{equation}\label{tp udefj ht}\times\biggl((\hat{p}^{\alpha}_{i}\psi)^{+}(R,t)\psi(R,t)+\psi^{+}(R,t)(\hat{p}^{\alpha}_{i}\psi)(R,t)\biggr).\end{equation}
The momentum balance equation for the system of particles under
consideration is obtained similarly, i.e., via differentiation of
current density (\ref{tp udefj ht}) and applying the
Schr\"{o}dinger equation ~\cite{Maksimov 99}, ~\cite{Maksimov 01},
~\cite{Andreev PRA08}. As a result, we obtain:
$$m\partial_{t}j^{\alpha}(\textbf{r},t)+\partial_{\beta
}\Pi^{\alpha\beta}(\textbf{r},t)$$
$$=-\int
d\textbf{r}'(\nabla^{\alpha}U(\textbf{r},\textbf{r}'))n_{2}(\textbf{r},\textbf{r}',t)$$
$$-\int d\textbf{r}'\int d\textbf{r}''(\partial^{\alpha}U(\textbf{r},\textbf{r}',\textbf{r}''))n_{3}(\textbf{r},\textbf{r}',\textbf{r}'',t)$$
\begin{equation}\label{tp Ebi}-n(\textbf{r},t)\nabla^{\alpha}V_{ext}(\textbf{r},t).\end{equation}
In the
momentum balance equation, $\Pi^{\alpha\beta}(\textbf{r},t)$ is
the quantum tensor of the density of the momentum flux. This
tensor has the form
$$\Pi^{\alpha\beta}(\textbf{r},t)=\int
 dR\sum_{i}\delta(\textbf{r}-\textbf{r}_{i})$$
$$\times\frac{1}{4m_{i}}\biggl(\psi^{+}(R,t)(\hat{p}^{\alpha}_{i}\hat{p}^{\beta}_{i}\psi)(R,t)$$
\begin{equation} \label{tp uPi ht}+(\hat{p}^{\alpha}_{i}\psi)^{+}(R,t)(\hat{p}^{\beta}_{i}\psi)(R,t)+c.c.\biggr). \end{equation}
The interaction between the particles in (\ref{tp Ebi}) is
expressed through the two-particle probability density
$n_{2}(\textbf{r},\textbf{r}',t)$ normalized over $N(N-1)$ and
having the form
$$n_{2}(\textbf{r},\textbf{r}',t)=\int dR\sum_{i,j\neq i}\delta(\textbf{r}-\textbf{r}_{i})$$
\begin{equation} \label{tp n2def}\times\delta(\textbf{r}'-\textbf{r}_{j})\psi^{+}(R,t)\psi(R,t) ,\end{equation}
and three-particle probability density
$n_{3}(\textbf{r},\textbf{r}',\textbf{r}'',t)$, normalized over
$N(N-1)(N-2)$ and having the form:
$$n_{3}(\textbf{r},\textbf{r}',\textbf{r}'',t)=\int dR\sum_{i,j\neq i;k\neq i,j}\delta(\textbf{r}-\textbf{r}_{i})$$
\begin{equation} \label{tp n3def}\times\delta(\textbf{r}'-\textbf{r}_{j})\delta(\textbf{r}''-\textbf{r}_{k})\psi^{+}(R,t)\psi(R,t).\end{equation}
 Let us represent the first term in the second
member of equation (\ref{tp Ebi}), i.e., the density of the
interaction force of the particles, in the form
$$-\frac{1}{2}\int dR\sum_{i,j.i\neq j}(\delta(\textbf{r}-\textbf{r}_{i})-\delta(\textbf{r}-\textbf{r}_{j}))$$
\begin{equation} \label{tp usigma1} \times(\nabla^{\alpha}_{i}U(\textbf{r}_{ij}))\psi^{+}(R,t)\psi(R,t),\end{equation}
 which is possible by
virtue of symmetry (antisymmetry) of the wave function, and let us
proceed in (\ref{tp usigma1}) to variables of the center of
gravity and variables of the relative distance of the particles:
 \begin{equation} \label{tp def} \begin{array}{ccc} \textbf{R}_{ij}=\frac{1}{2}(\textbf{r}_{i}+\textbf{r}_{j}) ,& \textbf{r}_{ij}=\textbf{r}_{i}-\textbf{r}_{j} \end{array}.\end{equation}
Rewrite the three-particle interaction potential in the form:
$$\partial_{i}^{\alpha}U(r_{ij},r_{jk},r_{ki})$$
\begin{equation} \label{tp}=-\partial_{j}^{\alpha}U(r_{ij},r_{jk},r_{ki})-\partial_{k}^{\alpha}U(r_{ij},r_{jk},r_{ki}).\end{equation}
Second term on the right-hand side eq. (\ref{tp Ebi}) is
represented in the form
$$-\frac{1}{9}\int dR\sum_{i,j,k,i\neq j;k\neq
i,j}\Biggl(\delta(\textbf{r}-\textbf{r}_{i})\partial_{i}^{\alpha}U(r_{ij},r_{jk},r_{ki})$$
$$+\delta(\textbf{r}-\textbf{r}_{j})\partial_{j}^{\alpha}U(r_{ij},r_{jk},r_{ki})$$
$$+\delta(\textbf{r}-\textbf{r}_{k})\partial_{k}^{\alpha}U(r_{ij},r_{jk},r_{ki})\Biggr)\psi^{+}(R,t)\psi(R,t).$$
Using (\ref{tp}) we obtain
$$-\frac{1}{9}\int
dR\sum_{i,j,k,i\neq j;k\neq
i,j}([\delta(\textbf{r}-\textbf{r}_{i})-\delta(\textbf{r}-\textbf{r}_{k})]\partial_{i}^{\alpha}U_{ijk}$$
\begin{equation} \label{tp int three pat 1}+[\delta(\textbf{r}-\textbf{r}_{j})-\delta(\textbf{r}-\textbf{r}_{k})]\partial_{j}^{\alpha}U_{ijk})\psi^{+}(R,t)\psi(R,t).\end{equation}
For the case of three particles we can use variables of center of
mass and relative motion too.

$$\begin{array}{ccc} \textbf{R}_{ijk}=\frac{1}{3}(\textbf{r}_{i}+\textbf{r}_{j}+\textbf{r}_{k}) ,& \textbf{r}_{ij}=\textbf{r}_{i}-\textbf{r}_{j},&\textbf{r}_{ik}=\textbf{r}_{i}-\textbf{r}_{k}  \end{array},$$
\begin{equation} \label{tp def}\textbf{r}_{jk}=\textbf{r}_{j}-\textbf{r}_{k}=\textbf{r}_{ik}-\textbf{r}_{ij}.\end{equation}
Since the interaction forces between the particles rapidly descend
at distances of the order of the interaction radius, small
$|r^{\alpha}_{ij}|$ give the main contribution to the integral
(\ref{tp usigma1}). Therefore, in expressions (\ref{tp usigma1})
and (\ref{tp int three pat 1}) we can replace the multipliers at
the interaction potential by their expansion in series by
$r^{\alpha}_{ij}$. We have concluded that the density of the
interaction force for bosons with a short-range interaction
potential can be represented in the form of divergence of the
tensor field $\partial_{\beta}\sigma^{\alpha\beta}(\textbf{r},t)$.
Here, $\sigma^{\alpha\beta}(\textbf{r},t)$ is the quantum stress
tensor due to inter-particle interaction.

Therefore, the momentum balance equation will take the form
$$\partial_{t}j^{\alpha}(\textbf{r},t)+\frac{1}{m}\partial_{\beta
 }(\Pi^{\alpha\beta}(\textbf{r},t)+\sigma^{\alpha\beta}(\textbf{r},t))$$
\begin{equation}\label{tp eq b j}=-\frac{1}{m}n(\textbf{r},t)\nabla^{\alpha}V_{ext}(\textbf{r})
.\end{equation}

 Let us now consider the tensor $\Pi^{\alpha\beta}(\textbf{r},t)$ and
isolate in it the contributions to the momentum flow density for
the convective and thermal motions and the purely quantum part.
For this purpose, let us introduce velocities by formulas
\begin{equation}\textbf{v}_{i}(R,t)=\frac{1}{m_{i}}\nabla_{i}S(R,t)
,\end{equation} where $\textbf{v}_{i}(R,t)$ is the velocity of the
i-th particle, while the $S(R,t)$-phase of the wave function is
$$\psi(R,t)=a(R,t) exp(\frac{\imath S(R,t)}{\hbar}).$$
  Velocity
field $ \textbf{v}(\textbf{r},t) $ is determined by the formula
\begin{equation}\label{tp curr}\textbf{j}(\textbf{r},t)=n(\textbf{r},t)\textbf{v}(\textbf{r},t).\end{equation}
 Then
$\textbf{u}_{i}(\textbf{r},R,t)=\textbf{v}_{i}(R,t)-\textbf{v}(\textbf{r},t)$
is the quantum analog of the velocity of thermal motion. Isolating
the explicitly thermal motion of the particles with velocities
$\textbf{u}_{i}$ and the motion with the velocity
$\textbf{v}(\textbf{r},t)$ in continuity and momentum balance
equations (\ref{tp eq b j}), we come to the following equations:
\begin{equation}\label{tp cont1}
\partial_{t}n(\textbf{r},t)+\nabla(n(\textbf{r},t)\textbf{v}(\textbf{r},t))=0 ,\end{equation}
$$ mn(\textbf{r},t)(\partial_{t}+\textbf{v}\nabla)v^{\alpha}(\textbf{r},t)+n(\textbf{r},t)\nabla^{\alpha}V_{ext}(\textbf{r},t)$$
\begin{equation}\label{tp eiler v1} +\partial_{\beta}(p^{\alpha\beta}(\textbf{r},t)+\sigma^{\alpha\beta}(\textbf{r},t)+T^{\alpha\beta}(\textbf{r},t))=0 .\end{equation}
In equation (\ref{tp eiler v1})
\begin{equation}\label{tp pressure} p^{\alpha\beta}(\textbf{r},t)=\int dR\sum_{i=1}^{N}\delta(\textbf{r}-\textbf{r}_{i})a^{2}(R,t)m_{i}u^{\alpha}_{i}u^{\beta}_{i} .\end{equation}
 This tensor tends
to zero along with equality to zero of velocities of thermal
motion $\textbf{u}_{i}$ of the particles. Therefore, it has the
meaning of the kinetic pressure.

The tensor
 $T^{\alpha\beta}(\textbf{r},t)$
is proportional to $\hbar^{2}$ and has a purely quantum origin.
For the system of numerous noninteracting particles, this tensor
is
$$
T^{\alpha\beta}(\textbf{r},t)=-\frac{\hbar^{2}}{4m}\Biggl(\partial^{\alpha}\partial^{\beta}n(\textbf{r},t)
$$
\begin{equation}\label{tp Bom2}-\frac{1}{n(\textbf{r},t)}(\partial^{\alpha}n(\textbf{r},t))(\partial^{\beta}n(\textbf{r},t))\Biggr).\end{equation}
Therefore, in this section we have obtained general form of QHD
equations for system of particles with short-range interaction.

\section{\label{sec:level1}III. Calculation of quantum stress tensor}

One have the aim to calculate quantum stress tensor
$\sigma^{\alpha\beta}(\textbf{r},t)$ it is necessary to write
explicit form of $\sigma^{\alpha\beta}(\textbf{r},t)$ through wave
functions.

The first terms of series in quantum stress tensor $\sigma^{\alpha\beta}(\textbf{r},t)$
gives the main contribution.
Writing this terms alone, we have
$$ \sigma^{\alpha\beta}(\textbf{r},t)=-\frac{1}{2}\int
dR\sum_{i,j.i\neq j}\delta(\textbf{r}-\textbf{R}_{ij})$$
$$\frac{r^{\alpha}_{ij}r^{\beta}_{ij}}{\mid\textbf{r}_{ij}\mid}\frac{\partial U(\textbf{r}_{ij})}{\partial\mid\textbf{r}_{ij}\mid}\psi^{+}(R,t)\psi(R,t)$$
$$-\frac{1}{9}\int dR\sum_{i;i\neq j;k\neq i,j}\Biggl((r_{ij}^{\beta}\partial_{j}^{\alpha}+r_{ik}^{\beta}\partial_{k}^{\alpha})U(r_{ij},r_{ik},\mid\textbf{r}_{ij}-\textbf{r}_{ik}\mid)\Biggr)$$
\begin{equation} \label{tp sigma in 1 or}\times\delta(\textbf{r}-\textbf{R}_{ijk})\psi^{+}(R,t)\psi(R,t).\end{equation}
Where
$$\psi(R,t)=\psi(...,R_{ijk},...,R_{ijk},...,t)$$
 in the term
describing binary interaction, similarly in the term describing
three particle interaction:
$$\psi(R,t)=\psi(...,R_{ijk},...,R_{ijk},...,R_{ijk},...,t).$$
Using symmetry properties both wave function of bosons and
potential of interaction between particles we can transform
formula (\ref{tp sigma in 1 or}) to the view:
$$
\sigma^{\alpha\beta}(\textbf{r},t)=-\frac{1}{2}Tr(n_{2}(\textbf{r},\textbf{r}',t))\int
d \textbf{r}\frac{r^{\alpha}r^{\beta}}{r}\frac{\partial
U(r)}{\partial r}$$
$$-\frac{1}{9}Tr (n_{3}(\textbf{r}
,\textbf{r}',\textbf{r}'',t))\times$$
\begin{widetext}
\begin{equation} \label{tp sigma in 1 or hz n2}\times\int
d\textbf{r}_{12}d\textbf{r}_{13}\Biggl(r_{12}^{\beta}\Biggl(\frac{\partial}{\partial
r_{2}^{\alpha}}(U(r_{12},r_{13},\mid\textbf{r}_{12}-\textbf{r}_{13}\mid)\Biggr)+r_{13}^{\beta}\frac{\partial}{\partial
r_{3}^{\alpha}}(U(r_{12},r_{23},\mid\textbf{r}_{12}-\textbf{r}_{13}\mid)\Biggr),\end{equation}
\end{widetext}
where
$$ Tr f(\textbf{r},\textbf{r}')=f(\textbf{r},\textbf{r}) ,$$
$$Tr f(\textbf{r},\textbf{r}',\textbf{r}'')=f(\textbf{r},\textbf{r},\textbf{r}).$$
It is evident that the first term in the second series of
$\sigma^{\alpha\beta}(\textbf{r},t)$ (related with TPI) for the
case of fermions equals to zero. For investigation the influence
of TPI on dynamics of fermions one needs to use the next term of
this series.

Using knowledge of decomposition formulas of wave function and relations of orthogonality, which are presented in appendix I,
 we obtain following expression
for two-particle concentration
\begin{equation}\label{tp n2 long r}
  n_2(\textbf{r},\textbf{r}',t)=n(\textbf{r},t)n(\textbf{r}',t)+|\rho(\textbf{r},\textbf{r}',t)|^{2}+\wp(\textbf{r},\textbf{r}',t)
,\end{equation}
 and for three-particle concentration
 \begin{widetext}
$$ n_3(\textbf{r},\textbf{r}',\textbf{r}'',t)=%n(\textbf{r},t)n(\textbf{r}',t)+|\rho(\textbf{r},\textbf{r}',t)|^{2}+\sum_{g}n_{g}(n_{g}-1)|\varphi_{g}(\textbf{r},t)|^{2}|\varphi_{g}(\textbf{r}',t)|^{2}
n(\textbf{r},t)n(\textbf{r}',t)n(\textbf{r}'',t)+n(\textbf{r}'',t)\mid\rho(\textbf{r},\textbf{r}',t)\mid^{2}+n(\textbf{r},t)\mid\rho(\textbf{r}',\textbf{r}'',t)\mid^{2}$$
$$+n(\textbf{r}',t)\mid\rho(\textbf{r},\textbf{r}'',t)\mid^{2}+[\rho(\textbf{r},\textbf{r}',t)\rho(\textbf{r}',\textbf{r}'',t)\rho(\textbf{r}'',\textbf{r},t)+c.c.]$$
$$+n(\textbf{r}',t)\wp(\textbf{r},\textbf{r}'',t)+n(\textbf{r}'',t)\wp(\textbf{r},\textbf{r}',t)+n(\textbf{r},t)\wp(\textbf{r}',\textbf{r}'',t)$$
$$+[\rho(\textbf{r}'',\textbf{r}',t)\sum_{g}n_{g}(n_{g}-1)\Biggl(\varphi_{g}(\textbf{r},t)\varphi_{g}(\textbf{r}'',t)\varphi_{g}^{*}(\textbf{r},t)\varphi_{g}^{*}(\textbf{r}',t)\Biggr)+c.c.]$$
\begin{equation}\label{tp n3 long r}+\sum_{g}n_{g}(n_{g}-1)(n_{g}-2)|\varphi_{g}(\textbf{r},t)|^{2}|\varphi_{g}(\textbf{r}',t)|^{2}|\varphi_{g}(\textbf{r}'',t)|^{2}.\end{equation}
\end{widetext}
In (\ref{tp n2 long r}), (\ref{tp n3 long r}) there is not
distinction functions describing behavior of BEC and of particles
in exited states.

Here
\begin{equation}\label{tp nvarphi}
n(\textbf{r},t)=\sum_{g}n_{g}\varphi_{g}^{*}(\textbf{r},t)\varphi_{g}(\textbf{r},t)
,\end{equation}
\begin{equation}\label{tp rhovarphi}\rho(\textbf{r},\textbf{r}',t)=\sum_{g}n_{g}\varphi_{g}^{*}(\textbf{r},t)\varphi_{g}(\textbf{r}',t),\end{equation}
\begin{equation}\label{tp def of DD
function}\wp(\textbf{r},\textbf{r}',t)=\sum_{g}n_{g}(n_{g}-1)|\varphi_{g}(\textbf{r},t)|^{2}|\varphi_{g}(\textbf{r}',t)|^{2},
\end{equation}
 where $\varphi_{g}(\textbf{r},t)$ are the
arbitrary single-particle wave functions.

The last two terms in formula (\ref{tp n2 long r}) represent part
of stress tensor arose via exchange interaction. Substituting
expression (\ref{tp n2 long r}) in (\ref{tp sigma in 1 or hz n2})
for quantum stress tensor of  bosons system, taking into account
$Tr\rho(\textbf{r},\textbf{r}',t)=n(\textbf{r},t)$, one obtains
formula
$$\sigma^{\alpha\beta}(\textbf{r},t)=-\frac{1}{2}\Upsilon\delta^{\alpha\beta}(2n^{2}(\textbf{r},t)+\wp(\textbf{r},t))$$
$$-\frac{2}{3}\chi^{\alpha\beta}\Biggl(6n^{3}(\textbf{r},t)+5n(\textbf{r},t)\wp(\textbf{r},t)
$$
\begin{equation}\label{tp sigma boz}+\sum_{g}n_{g}(n_{g}-1)(n_{g}-2)\mid\varphi_{g}(\textbf{r},t)\mid^{6}\Biggr),\end{equation}
where
\begin{equation}\label{tp def of D
function}\wp(\textbf{r},t)=\sum_{g}n_{g}(n_{g}-1)|\varphi_{g}(\textbf{r},t)|^{4},\end{equation}
$$\wp(\textbf{r},t)=\wp_{B}(\textbf{r},t)+\wp_{n}(\textbf{r},t)$$
and
$$Tr\wp(\textbf{r},\textbf{r}',t)=\wp(\textbf{r},t).$$
 In the (\ref{tp sigma boz}) following integrals is designated through $\Upsilon$ and $\chi^{\alpha\beta}$
\begin{equation}\label{tp Upsilon} \Upsilon=\frac{4\pi}{3}\int
dr(r)^{3}\frac{\partial U(r)}{\partial r},
\end{equation}
\begin{widetext}
$$\chi^{\alpha\beta}\equiv-\frac{1}{6}\int d\textbf{r}_{1}d\textbf{r}_{2}\Biggl(\Biggl(\frac{r_{1}^{\alpha}r_{1}^{\beta}}{r_{1}}\partial_{1}+\frac{r_{2}^{\alpha}r_{2}^{\beta}}{r_{2}}\partial_{2}+2\frac{(r_{1}^{\alpha}-r_{2}^{\alpha})(r_{1}^{\beta}-r_{2}^{\beta})}{\mid \textbf{r}_{1}-\textbf{r}_{2}\mid}\partial_{3}\Biggr)$$
\begin{equation}\label{tp chi}\times U(r_{1},r_{2},\sqrt{r_{1}^{2}+r_{2}^{2}+2r_{1}r_{2}\cos\Omega})  \Biggr),\end{equation}
\end{widetext} where $\Omega$ is angle between  $\textbf{r}_{1}$
and $\textbf{r}_{2}$ and $\partial_{1}$, $\partial_{2}$,
$\partial_{3}$ are derivatives of function $U$ on its arguments.
We can see that $\chi^{\alpha\beta}=\chi^{\beta\alpha}$.

 Assuming that the potential
satisfies the condition that the quantity $r^{3}U(r)$ tends to
zero at $r$ tending to zero and infinity, for $\Upsilon$ from
(\ref{tp Upsilon}), by integration by parts, we obtain
$$ \Upsilon=-\int d\textbf{r}U(r) ,$$
which coincides with the result for the interaction constant $g$ found
by Gross and Pitaevskii allowing for the sign in  (\ref{tp eiler
v2}) $\Upsilon=-g$.

Taking into account the notions given after formula (\ref{tp
cap2:permanent}), we obtain the following relation for the quantum
tensor of stress of the system of bosons close to the BEC state:
$$\sigma_{B,n}^{\alpha\beta}(\textbf{r},t)=-\frac{1}{2}\Upsilon\delta^{\alpha\beta}$$
$$\times\Biggl(2n_{B}(\textbf{r},t)n_{n}(\textbf{r},t)+2n_{n}^{2}(\textbf{r},t)+\wp(\textbf{r},t)\Biggr)$$
$$-\frac{2}{3}\chi^{\alpha\beta}\Biggl(18n_{B}(\textbf{r},t)n_{n}^{2}(\textbf{r},t)+5n_{n}(\textbf{r},t)\wp(\textbf{r},t)$$
\begin{equation}\label{tp sigma boz and sm exitation}+6n_{n}^{3}(\textbf{r},t)+5n_{B}(\textbf{r},t)\wp_{n}(\textbf{r},t)+\widetilde{m}(\textbf{r},t)\Biggr),\end{equation}
where
$$\widetilde{m}(\textbf{r},t)=\sum_{g}n_{g}(n_{g}-1)(n_{g}-2)\mid\varphi_{g}(\textbf{r},t)\mid^{6}.$$
Here, we used the notations $n_{B}(\textbf{r},t)$ for the
concentration of particles situating in the BEC state and
$n_{n}(\textbf{r},t)$ for the concentration of excited particles.
Notation $g_{0}$ designates ground state of system of particles
corresponding to BEC.

 The
stress tensor that depends on inter-particle interactions contains
the single-particle functions $\varphi_{g}(\textbf{r},t)$. By
these functions, expansion of the unknown N-particles wave
function is actually performed. If we neglect the inter-particle
interaction, the  N-particle problem will be reduced to the
single-particle one, and the set of functions
$\varphi_{g}(\textbf{r},t)$ will be determined by the
single-particle Schr\"{o}dinger equation. In this case, the
momentum balance equation will also not contain interactions and,
along with the continuity equation, will determine the
single-particle wave function in the polar form. Such
single-particle functions, which are simultaneously solutions of
equations of quantum hydrodynamics of the system of noninteracting
particles and the single-particle Schr\"{o}dinger equation, can be
used as the first approximation at calculating of the stress
tensor. Thus obtained quantum balance equations will determine the
set of field functions that determine the state of the system in
hydrodynamics and, consequently, the single-particle wave function
of the system of interacting particles. Such a wave function of
spatial coordinates and time will determine the effective
Schr\"{o}dinger equation, which is nonlinear and
integro-differential in the general case. The solution of such
equation can be used as the second iteration at calculating the
stress tensor, etc.

 For particles to be found in
state of Bose condensation we can get relation for $\wp_{B}$ and $\widetilde{m}_{B}$
$$\wp_{B}(\textbf{r},t)=\sum_{g}n_{g}(n_{g}-1)|\varphi_{g}(\textbf{r},t)|^{4}$$
$$=\sum_{g}n_{g}^{2}|\varphi_{g}(\textbf{r},t)|^{4}=N^{2}|\varphi_{g_{0}}(\textbf{r},t)|^{4}$$
\begin{equation}\label{tp corr.n2}=(N|\varphi_{g_{0}}(\textbf{r},t)|^{2})^{2}=n_{B}^{2}(\textbf{r},t),\end{equation}
similarly:
$$\widetilde{m}_{B}(\textbf{r},t)=\sum_{g}n_{g}(n_{g}-1)(n_{g}-2)\times$$
\begin{equation}\label{tp corr.n3}\times|\varphi_{g}(\textbf{r},t)|^{6}=n_{B}^{3}(\textbf{r},t).\end{equation}

Using results (\ref{tp corr.n2}), (\ref{tp corr.n3}), we obtain
expression for $\sigma_{B,n}^{\alpha\beta}(\textbf{r},t)$ in the
form:
$$\sigma_{B,n}^{\alpha\beta}(\textbf{r},t)=-\frac{1}{2}\Upsilon\delta^{\alpha\beta}\biggl(2n_{B}(\textbf{r},t)n_{n}(\textbf{r},t)$$
$$+n^{2}_{B}(\textbf{r},t)+2n_{n}^{2}(\textbf{r},t)+\wp_{n}(\textbf{r},t)\biggr)-\frac{2}{3}\chi^{\alpha\beta}\Biggl(6n_{n}^{3}(\textbf{r},t)$$
$$+18n_{B}(\textbf{r},t)n_{n}^{2}(\textbf{r},t)+5n_{n}(\textbf{r},t)n^{2}_{B}(\textbf{r},t)$$
\begin{equation}\label{tp sigma boz and sm exitation}+5n_{B}(\textbf{r},t)\wp_{n}(\textbf{r},t)+n_{B}^{3}(\textbf{r},t)+\widetilde{m}_{n}\Biggr).\end{equation}
Without exited particles quantum stress tensor takes the form:
\begin{equation}\label{tp sigma boz only BEC}\sigma_{B}^{\alpha\beta}(\textbf{r},t)=-\frac{1}{2}\Upsilon\delta^{\alpha\beta}n^{2}_{B}(\textbf{r},t)-\frac{2}{3}\chi^{\alpha\beta} n_{B}^{3}(\textbf{r},t).\end{equation}
The corresponding momentum balance equation of the quantum
hydrodynamics for BEC alone takes the form:
$$ mn(\textbf{r},t)(\partial_{t}v^{\alpha}(\textbf{r},t)+v^{\beta}(\textbf{r},t)\nabla^{\beta}v^{\alpha}(\textbf{r},t))+\partial^{\beta}p^{\alpha\beta}(\textbf{r},t)
$$
$$-\frac{\hbar^{2}}{2m}n(\textbf{r},t)\partial_{\alpha}\frac{\triangle\sqrt{n(\textbf{r},t)}}{\sqrt{n(\textbf{r},t)}}-\Upsilon n(\textbf{r},t)\partial^{\alpha}n(\textbf{r},t)$$
\begin{equation}\label{tp eiler v2}-2\chi^{\alpha\beta} n^{2}(\textbf{r},t)\partial^{\beta}n(\textbf{r},t)=-
n(\textbf{r},t)\nabla^{\alpha}V_{ext}(\textbf{r},t).
\end{equation}
Therefore in this section we have obtained the QHD equations for
BEC with two- and three-particle interaction.

\section{\label{sec:level1}IV. Eiler equations for a boson system with nonzero temperature}

If we interesting in separate dynamics of BEC and temperature excited
bosons we can divide the concentration $n(\textbf{r},t)$ and current $\textbf{j}(\textbf{r},t)$ on two part
$$n(\textbf{r},t)=n_{B}(\textbf{r},t)+n_{n}(\textbf{r},t)$$
and
$$\textbf{j}(\textbf{r},t)=\textbf{j}_{B}(\textbf{r},t)+\textbf{j}_{n}(\textbf{r},t).$$

Here we consider a case there are no exchange of particles between
BEC and noncondensate component. In this situation we can write
the continuity equations for each kinds of particles, for the BEC
\begin{equation}\label{tp cont B}\partial_{t}n_{B}(\textbf{r},t)+\nabla(n_{B}(\textbf{r},t)\textbf{v}_{B}(\textbf{r},t))=0\end{equation}
and, for the noncondensate bosons
\begin{equation}\label{tp cont n}\partial_{t}n_{n}(\textbf{r},t)+\nabla(n_{n}(\textbf{r},t)\textbf{v}_{n}(\textbf{r},t))=0.\end{equation}
In previous section we derive the momentum balance equation for
whole system of bosons, i.e. for the mixture of BEC and
noncondensate particles. Now we need to obtain momentum balance
equations for each kinds of bosons. For separation of
contributions of different sorts of bosons we must to consider of
derivation of quantum stress tensor (\ref{tp n2 long r}) and
(\ref{tp n3 long r}) \textit{or} (\ref{tp sigma boz and sm
exitation}), in more detailed way.

For understanding the detail of evolution of
$\textbf{j}_{B}(\textbf{r},t)$ and $\textbf{j}_{n}(\textbf{r},t)$
we need to consider formulas (\ref{tp cap2:permanent}) and
(\ref{tp cap2:permanent 4}) in detail. The first multiplier in
this formula, which has argument $(\textbf{r},t)$ is related to
the particle whose motion we consider. Another one particle wave
functions are related to the particles that influence on dynamic
of considered current. This is give us ability to obtain the
separate equation of dynamic atoms in BEC state and noncondensate
ones. We consider each term in (\ref{tp cap2:permanent}) and
(\ref{tp cap2:permanent 4}). If one-particle wave function with
argument $(\textbf{r},t)$ describe BEC state (has subindex "B"),
we put this term in momentum balance equation for BEC. In the case
one-particle wave function with argument $(\textbf{r},t)$ describe
noncondensed state we put this term in momentum balance equation
for noncondesate particles. In this way we get following
expressions for quantum stress tensors of atoms in both BEC and
noncondensed states.

$$\sigma^{\alpha\beta}_{B}=-\frac{1}{2}\Upsilon\delta^{\alpha\beta}(n_{B}n_{n}+\wp_{B})$$
\begin{equation}\label{tp moment with temp BEC}-\frac{2}{3}\chi^{\alpha\beta}(6n_{B}n_{n}^{2}+4n_{n}\wp_{B}+n_{B}\wp_{n}+\hat{m}_{B})\end{equation}
and
$$\sigma^{\alpha\beta}_{n}=-\frac{1}{2}\Upsilon\delta^{\alpha\beta}(n_{B}n_{n}+2n_{n}^{2}+\wp_{n})$$
\begin{equation}\label{tp moment with temp norm}-\frac{2}{3}\chi^{\alpha\beta}(6n_{n}^{3}+12n_{B}n_{n}^{2}+5n_{n}\wp_{n}+4n_{B}\wp_{n}+n_{n}\wp_{B}+\hat{m}_{n}).\end{equation}

In appendix 2 we calculate $\wp_{n}$ and $\hat{m}_{n}$ and receive
$$\begin{array}{ccc}\wp_{n}=0,& \hat{m}_{n}=0.&\end{array}$$

From formulas (\ref{tp corr.n2}) and (\ref{tp corr.n3}) we see
$\wp_{B}(\textbf{r},t)=n_{B}^{2}(\textbf{r},t)$ and
$\widetilde{m}_{B}(\textbf{r},t)=n_{B}^{3}(\textbf{r},t)$. In
this case we have $\sigma^{\alpha\beta}_{B}$ and
$\sigma^{\alpha\beta}_{n}$ has more simple form and expressed in
terms of $n_{B}$ and $n_{n}$.

Obtained expression (\ref{tp sigma boz}) for
$\sigma^{\alpha\beta}(\textbf{r},t)$
in the approximation under
consideration should be substituted into momentum balance equation
(\ref{tp eiler v1}).

Using this results and formulas (\ref{tp corr.n2}), (\ref{tp
corr.n3}) we can rewrite equations (\ref{tp moment with temp BEC})
and (\ref{tp moment with temp norm}) in the form
$$mn_{B}(\textbf{r},t)(\partial_{t}v^{\alpha}_{B}(\textbf{r},t)+v^{\beta}_{B}(\textbf{r},t)\nabla^{\beta}v^{\alpha}_{B}(\textbf{r},t))$$
$$-\frac{\hbar^{2}}{2m}n_{B}(\textbf{r},t)\partial_{\alpha}\frac{\triangle\sqrt{n_{B}(\textbf{r},t)}}{\sqrt{n_{B}(\textbf{r},t)}}$$
$$=-n_{B}(\textbf{r},t)\nabla^{\alpha}V_{ext}(\textbf{r},t)+\frac{1}{2}\Upsilon\partial^{\alpha}(n_{B}n_{n}+n_{BEC}^{2})$$
\begin{equation}\label{tp moment with temp with corr BEC}+\frac{2}{3}\chi^{\alpha\beta}\partial^{\beta}\biggl(6n_{B}n_{n}^{2}+4n_{n}n_{BEC}^{2}+n_{BEC}^{3}\biggr)\end{equation}
and
$$mn_{n}(\textbf{r},t)(\partial_{t}v^{\alpha}_{n}(\textbf{r},t)+v^{\beta}_{n}(\textbf{r},t)\nabla^{\beta}v^{\alpha}_{n}(\textbf{r},t))$$
$$-\frac{\hbar^{2}}{2m}n_{n}(\textbf{r},t)\partial_{\alpha}\frac{\triangle\sqrt{n_{n}(\textbf{r},t)}}{\sqrt{n_{n}(\textbf{r},t)}}$$
$$=-n_{n}(\textbf{r},t)\nabla^{\alpha}V_{ext}(\textbf{r},t)+\frac{1}{2}\Upsilon\partial^{\alpha}(n_{B}n_{n}+2n_{n}^{2})$$
\begin{equation}\label{tp moment with temp with corr norm}+\frac{2}{3}\chi^{\alpha\beta}\partial^{\beta}\biggl(6n_{n}^{3}+12n_{B}n_{n}^{2}+n_{n}n_{BEC}^{2}\biggr).\end{equation}
where evident form of interaction constants
$\Upsilon$ and $\chi^{\alpha\beta}$ are presented
by formulas (\ref{tp Upsilon}) and (\ref{tp chi}) correspondingly.

\section{\label{sec:level1}V. Derivation of NLS equation for zero temperature}

There is used the NLSE along with QHD equations in the literature.
From QHD equations one can obtain the NLSE ~\cite{Andreev PRA08,Maksimov 99}.
Macroscopic wave function may be defined
via hydrodynamic variables: the concentration of particle number
n(\textbf{r},t) and potential of velocities field
$\phi(\textbf{r},t)$. For determination of evolution of velocity
field the integral of Cauchy-Lagrange following of the momentum
balance equation is used.

Therefore, for eddy-free motion
$\textbf{v}(\textbf{r},t)=\nabla\phi(\textbf{r},t)$ and
barotropicity condition in terms of tensor field
\begin{equation}\label{tp barotropnost tensor}\frac{\partial^{\beta} p^{\alpha\beta}(\textbf{r},t)}{mn(\textbf{r},t)}=\partial^{\beta} \mu^{\alpha\beta}(\textbf{r},t),\end{equation}
where $\mu^{\alpha\beta}(\textbf{r},t)$-tensor of chemical
potential, momentum balance equation
has the first tensor integral:
$$\delta^{\alpha\beta}\partial_{t}\phi(\textbf{r},t)+\frac{1}{2}\delta^{\alpha\beta}v^{2}(\textbf{r},t)+\mu^{\alpha\beta}(\textbf{r},t)$$
$$-\frac{1}{m}\Upsilon\delta^{\alpha\beta} n(\textbf{r},t)-\frac{1}{m}\chi^{\alpha\beta} n^{2}(\textbf{r},t)+\frac{1}{m}V_{ext}(\textbf{r},t)\delta^{\alpha\beta}$$
\begin{equation}\label{tp int Koshy tensor}-\frac{\hbar^{2}}{2m^{2}}\delta^{\alpha\beta}\frac{\triangle\sqrt{n(\textbf{r},t)}}{\sqrt{n(\textbf{r},t)}}=(const)^{\alpha\beta}.\end{equation}
Continuity equation (\ref{tp cont1}) and momentum balance equation
(\ref{tp eiler v2}) can be associated with the equivalent
one-particle Schr\"{o}dinger equation for some effective wave
function $\Phi(\textbf{r},t)$. Represent this function in the
form:
\begin{equation}\label{tp WF in m}
\Phi(\textbf{r},t)=\sqrt{n(\textbf{r},t)}\exp\biggl(\frac{\imath}{\hbar}m\phi(\textbf{r},t)\biggr)
.\end{equation}
Differentiating it by time and using equations
(\ref{tp cont1}), (\ref{tp int Koshy tensor}) we take equivalent
one-particle Schr\"{o}dinger equation:
$$\imath\hbar\delta^{\alpha\beta}\partial_{t}\Phi(\textbf{r},t)=\Biggl(-\delta^{\alpha\beta}\frac{\hbar^{2}\nabla^{2}}{2m}+\mu^{\alpha\beta}(\textbf{r},t)+V_{ext}(\textbf{r},t)\delta^{\alpha\beta}$$
\begin{equation}\label{tp u GP tensor} -\Upsilon\delta^{\alpha\beta}\mid\Phi(\textbf{r},t)\mid^{2}-\chi^{\alpha\beta}\mid\Phi(\textbf{r},t)\mid^{4}\Biggr)\Phi(\textbf{r},t)
,\end{equation} under conditions
$\chi^{\alpha\beta}=\chi\delta^{\alpha\beta}$,
$p^{\alpha\beta}(\textbf{r},t)=p(\textbf{r},t)\delta^{\alpha\beta}$
isotropic kinetic pressure and barotropicity condition:
\begin{equation}\label{tp barotropnost}\frac{\nabla p(\textbf{r},t)}{mn(\textbf{r},t)}=\nabla \mu(\textbf{r},t),\end{equation}
where $\mu(\textbf{r},t)$-chemical potential, we obtain:
$$\imath\hbar\partial_{t}\Phi(\textbf{r},t)=\Biggl(-\frac{\hbar^{2}\nabla^{2}}{2m}+\mu(\textbf{r},t)$$
\begin{equation}\label{tp u GP} +V_{ext}(\textbf{r},t)-\Upsilon\mid\Phi(\textbf{r},t)\mid^{2}-\chi\mid\Phi(\textbf{r},t)\mid^{4}\Biggr)\Phi(\textbf{r},t)
.\end{equation}
That is well known Gross-Pitaevskii equation
~\cite{L.P.Pitaevskii RMP 99,Gross 61,L.P.Pitaevskii 61,Gross 63},
 with the nonlinearity
of fifth degree ~\cite{Kovalev FNT 76,Barashenkov PLA
88}, which arise due to TPI.

The isotropic constant of three-particle interaction is:
$$\chi\equiv-\frac{1}{3}\int d\textbf{r}_{1}d\textbf{r}_{2}\Biggl(\Biggl(r_{1}\partial_{r_{1}}+\mid\textbf{r}_{1}-\textbf{r}_{2}\mid\partial_{3}\Biggr)$$
\begin{equation}\label{tp chi isotrop}\times U(r_{1},r_{2},\sqrt{r_{1}^{2}+r_{2}^{2}+2r_{1}r_{2}\cos\Omega})  \Biggr).\end{equation}

Wave function $\Phi(\textbf{r},t)$ is normalized with condition:
$$ \int d\textbf{r}\Phi(\textbf{r},t)^{*}\Phi(\textbf{r},t)=N ,$$
where $N$ is number of particles in the system.

Within this approximation the equation of momentum balance has
form:
$$
m\partial_{t}v^{\alpha}(\textbf{r},t)+\frac{1}{2}m\partial^{\alpha}v^{2}(\textbf{r},t)+m\partial^{\alpha}\mu(\textbf{r},t)$$
$$-\frac{\hbar^{2}}{2m}\partial^{\alpha}\frac{\triangle\sqrt{n(\textbf{r},t)}}{\sqrt{n(\textbf{r},t)}}-\Upsilon\partial^{\alpha}n(\textbf{r},t)$$
\begin{equation}\label{tp eiler TPI scal}-\chi\partial^{\alpha}n^{2}(\textbf{r},t)=-\partial^{\alpha}V_{ext}(\textbf{r},t).\end{equation}

If system of particles is dense (e.g. Bose liquid)  then two more
particles can interaction simultaneously. Basic contribution in
interaction three and more particles arise from binary
interactions $U_{ij}$. Part of interactions in macroscopic system
of particles which is described via binary potential is the origin
of first series in (\ref{tp sigma in 1 or}) ~\cite{Andreev PRA08}
and, consequently, this part is also the origin of the term which
is proportional $\Upsilon$ in (\ref{tp u GP}).

\section{\label{sec:level1}VI. Linear wave dispersion in BEC with TPI zero temperature limit}

In this section we have the aim both to consider dispersion
properties of linear wave and to compare two corrections to GP
approximation. For this aim we write system of quantum
hydrodynamic equations in third order  to the interaction radius,
for only binary interaction which was given in the work
~\cite{Andreev PRA08}. Taking in account binary
and three-particle interaction and including results of the
previous section we obtain equation:
$$mn(\textbf{r},t)\partial_{t}v^{\alpha}(\textbf{r},t)+\frac{1}{2}mn(\textbf{r},t)\nabla^{\alpha}v^{2}(\textbf{r},t)+\nabla^{\alpha}\mu(\textbf{r},t)$$
$$ -\frac{\hbar^{2}}{2m}n(\textbf{r},t)\partial^{\alpha}\frac{\triangle\sqrt{n(\textbf{r},t)}}{\sqrt{n(\textbf{r},t)}}-\frac{1}{2}\Upsilon\partial_{\alpha
 }n^{2}(\textbf{r},t)-\chi\partial^{\alpha}n^{3}(\textbf{r},t)$$
 \begin{equation}\label{tp eiler boze TOIR and TPI} -\frac{1}{2}\Upsilon_{2}\partial_{\alpha}\triangle n^{2}(\textbf{r},t)=-n(\textbf{r},t)\nabla^{\alpha}V_{ext}(\textbf{r},t).\end{equation}
where
$$\Upsilon_{2}=\frac{\pi}{30}\int dr
(r)^{5}\frac{\partial U(r)}{\partial r}.$$
Following to the article ~\cite{Andreev arxiv 11 2} we redefined $\Upsilon_{2}$, in compare with the paper ~\cite{Andreev PRA08}.

 Let us consider the eigenmodes, which can propagate in one
dimensional geometry,  based on the  Eqs. of continuity (\ref{tp
cont1}) and momentum balance (\ref{tp eiler boze TOIR and TPI}).

We consider the small perturbation of equilibrium state like
\begin{equation}\label{tp equlib state BEC}\begin{array}{ccc}n=n_{0}+\delta n,& v^{\alpha}=0+v^{\alpha},&\end{array}\end{equation}
Substituting these relations into system of equations  (\ref{tp
cont1}) and (\ref{tp equlib state BEC}) and neglecting nonlinear
terms, we obtain a system of linear homogeneous equations in
partial derivatives with constant coefficients. Passing to the
following representation for small perturbations $\delta f$
\begin{equation}\label{tp FFF}\delta f =f(\omega, \textbf{k}) exp(-\imath\omega+\imath \textbf{k}\textbf{r}) \end{equation}
yields the homogeneous system of algebraic equations. The
magnitude of concentration of BEC is assumed to have a nonzero
value. Expressing all the quantities entering the system of
equations in terms of the concentration of BEC, we come to the
dispersion equation for elementary excitations
\begin{equation}\label{tp dispersion General} \omega^{2}=\Biggl(\frac{\hbar^{2}}{4m^{2}}+\frac{n_{0}\Upsilon_{2}}{m}\Biggr)k^{4}-\Biggl(\frac{\Upsilon n_{0}}{m}+\frac{2\chi}{m}n_{0}^{2}\Biggr)k^{2} .\end{equation}

In the absence of $\Upsilon_{2}$ the result (\ref{tp dispersion General}) is in  accordance with the
real part of solution $\omega(k)$ obtained in ~\cite{Marklund
EPJ.B. 05}. If we compare terms which proportional to $\chi$ and $\Upsilon_{2}$ we can see functional dependence $\omega(k)$
for the cases three-particle interaction and binary interaction in
third order on interaction radius. When
we calculate binary interaction in third order to interaction
radius there arise linear dependence on $n_{0}$. Term proportional
to $k^{2}$ in (\ref{tp dispersion General}), las in GP approximation,
is linear on concentration $n_{0}$.  There is additional
dependence on $n_{0}^{2}$ in (\ref{tp dispersion General}) at the
expense of three-particle interaction.

\section{\label{sec:level1}VII. Linear wave dispersion in BEC with TPI nonzero temperature limit}

In this section we consider a Bose particle system contained
particles being in the BEC state  and noncondensate atoms. We
interested the dispersion  of small amplitude elementary
excitation in such system. For this aim we use system of equations
(\ref{tp cont n}), (\ref{tp cont B}), (\ref{tp moment with temp
with corr BEC}) and (\ref{tp moment with temp with corr norm}).
Equilibrium state is described by constant values of
concentrations of each sorts of bosons $n_{0B}$ and $n_{0n}$. The
flows in equilibrium state are absent
$\textbf{v}_{B}=\textbf{v}_{n}=0$. For small perturbations of
equilibrium state we have
\begin{equation}\label{tp equlib state BEC temper}\begin{array}{ccc}n_{i}=n_{0i}+\delta n_{i},& v^{\alpha}_{i}=0+v^{\alpha}_{i},&\end{array}\end{equation}
where $i$-mark $B$ or $n$. Using formula (\ref{tp FFF}) for small
perturbations we get the homogeneous system of algebraic
equations. A condition of existence on nontrivial solution lead to
dispersion equation. A solution of dispersion equation has form

\begin{widetext}
$$\omega^{2}=\frac{\hbar^{2}k^{4}}{4m^{2}}-\frac{\Upsilon k^{2}}{4m}(3n_{0B}+5n_{0n})-\frac{4}{3m}k^{\alpha}k^{\beta}\chi^{\alpha\beta}(n_{0B}^{2}+6n_{0n}^{2}+8n_{0B}n_{0n})$$
$$\pm\frac{1}{2m}\biggl(\frac{\Upsilon^{2}k^{4}}{4}(n_{0B}^{2}+9n_{0n}^{2}-2n_{0B}n_{0n})
+\frac{16}{9}(k^{\alpha}k^{\beta}\chi^{\alpha\beta})^{2}(n_{0B}^{4}+36n_{0n}^{4}
+124n_{0B}^{2}n_{0n}^{2}+240n_{0B}n_{0n}^{3}-8n_{0B}^{3}n_{0n})$$
\begin{equation}\label{tp disp general non zero temp}+\frac{4}{3}\Upsilon k^{2}k^{\alpha}k^{\beta}\chi^{\alpha\beta}(n_{0B}^{3}+18n_{0n}^{3}-5n_{0B}^{2}n_{0n}+42n_{0B}n_{0n}^{2})\biggr)^{1/2}.\end{equation}
\end{widetext}
In the case $\chi^{\alpha\beta}=\chi\delta^{\alpha\beta}$ we
have $k^{\alpha}k^{\beta}\chi^{\alpha\beta}=k^{2}\chi$ and we
can make $k^{2}$ out of the square root. Then, all terms, apart from first
term, are proportional to $k^{2}$. A coefficient at $k^{2}$ it
is a square of velocity of sound. Due to two sign at the square root
we have two sound velocity, for the low temperature first and second sounds.

The first sound it is a usual sound, one take place in classic
gases and in Bose system far of BEC condition. For the low
temperature quantum velocity of first sound coincide sign "-" in
the equation (\ref{tp disp general non zero temp}). Sign "+" in
equation (\ref{tp disp general non zero temp}) coincide to the
second sound it is exist in superfluid systems due to BEC.
Moreover, solution with sign "+" in equation (\ref{tp disp general
non zero temp}) it is the Bogoliubov's mode in BEC. In comparison
with Bogoliubov's, our result account three particle interaction
and influence of noncondensate particles.

In the absence of TPI $k^{\alpha}k^{\beta}\chi^{\alpha\beta}=0$,
for repulsive two-particle interaction $\Upsilon<0$, from (\ref{tp
disp general non zero temp}) we have
$$\omega^{2}=\frac{\hbar^{2}k^{4}}{4m^{2}}+\frac{\mid\Upsilon\mid k^{2}}{4m}\times\biggl(3n_{0B}+5n_{0n}$$
\begin{equation}\label{tp }\pm\sqrt{n_{0B}^{2}+9n_{0n}^{2}-2n_{0B}n_{0n}}\biggr)\end{equation}
where the quantity in bracket is positive.

For the case $\Upsilon=0$ and $k^{\alpha}k^{\beta}\chi^{\alpha\beta}<0$ we obtain
\begin{widetext}
$$\omega^{2}=\frac{\hbar^{2}k^{4}}{4m^{2}}+\frac{k^{2}\mid k^{\alpha}k^{\beta}\chi^{\alpha\beta}\mid}{3m}\biggl(2(n_{0B}^{2}+6n_{0n}^{2}+8n_{0B}n_{0n})$$
\begin{equation}\label{tp }\pm\sqrt{n_{0B}^{4}+36n_{0n}^{4}+124n_{0B}^{2}n_{0n}^{2}+240n_{0B}n_{0n}^{3}-8n_{0B}^{3}n_{0n}}\biggr)\end{equation}
\end{widetext} where the quantity in bracket is positive.

\section{\label{sec:level1}VIII. Conclusion}

In this article we gave derivation of the NLSE for BEC with cubic
and quintic nonlinearities from microscopic quantum theory. For
the derivation we used method of quantum hydrodynamics. In
original Schr\"{o}dinger equation we took into account
two-particle and three-particle interactions. In the article,
 arising force field of whole accounted interaction in the form of
divergence of tensor field, is demonstrated. We obtained the
expression of quantum stress tensor by means of microscopic wave
function. Using its representation we derived equation of state
for boson systems in BEC state including two- and three-particle
interaction. Therefore, dependence of quantum stress tensor
$\sigma^{\alpha\beta}(\textbf{r},t)$ on concentration
$n(\textbf{r},t)$ are demonstrated. Thus, momentum balance
equation obtained coincides with the analogous NLSE containing
nonlinearities of third and fifth degrees. Method of finding of
NLSE for the wave function in the medium for the case
three-particle interaction are developed. We showed that obtained
NLSE coincides with well-known NLSE with cubic and quintic
nonlinearities. Explicit form for the constant of three-particle
interaction was obtained. In particular, here represent derivation
of GP equation from many-particle Schr\"{o}dinger equation. Tensor
form of constant of three-particle interaction  is shown.

From derivation of QHD equations including three particle
interaction, obtained in this article, and from derivation  QHD
equations including TOIR, obtained in ~\cite{Andreev PRA08}, we
can see terms due to TOIR and TPI additional.

 In this article, frequency dependence of elementary excitation on wave
vector was calculated for the case including FOIR, TOIR and three
particle interaction. Comparison of contributions from TOIR and
three particle interaction is discussed.

A special attention we got for influence of the temperature on
dynamic of Bose particles. We made a microscopic derivation of
two-fluid hydrodynamic equation. We confine oneself by  continuity
equations and momentum balance equations for each type of bosons,
i.e. particles in BEC state and noncondensate particles, due to
two- and three-particle interaction, in first order on interaction
radius. In described approximation we studied a spectrum of
elementary excitations. We obtain the dispersion for two waves.
One of them it is generalization of Bogoliubov's mode, another it
is low temperature approximation for the usual sound.

\section{\label{sec:level1}ACKNOWLEDGMENTS}
The author wish to thank L.S. Kuz'menkov for discussion of the results obtained.

\section{\label{sec:level1}Appendix 1}

For two-particle density of probability  in correspondence with
the definition (\ref{tp n2def}) we have:
$$n_2(\textbf{r},\textbf{r}',t)$$
 $$
 =N(N-1)
  \int dR_{N-2}\langle n_1,n_2,\ldots |\textbf{r},\textbf{r}',R_{N-2},t\rangle$$
\begin{equation}\label{tp cap2:def.N_2}\times
  \langle\textbf{r},\textbf{r}',R_{N-2},t |n_1,n_2,\ldots\rangle
  ,\end{equation}
where $dR_{N-2}=\displaystyle\prod\limits_{k=3}^{N}d\textbf{r}_k$.

For three-particle density of probability we have at the same way:
$$ n_3(\textbf{r},\textbf{r}',\textbf{r}'',t)=N(N-1)(N-2)\times$$
$$
  \times\int dR_{N-3}\langle n_1,n_2,\ldots |\textbf{r},\textbf{r}',\textbf{r}'',R_{N-3},t\rangle
  \times$$
\begin{equation}\label{tp cap2:def.N_3}\times
  \langle\textbf{r},\textbf{r}',\textbf{r}'',R_{N-3},t |n_1,n_2,\ldots\rangle
  ,\end{equation}
where $dR_{N-3}=\displaystyle\prod\limits_{k=4}^{N}d\textbf{r}_k$.

Using knowledge of decomposition formulas  of wave function
$\langle\textbf{r},\textbf{r}',R_{N-2},t |n_1,n_2\ldots\rangle$ in
formulas (\ref{tp corr.n2}) and (\ref{tp corr.n3}) we use
following decomposition formulas ~\cite{Shveber}
\begin{widetext}
$$  \langle \textbf{r}, \textbf{r}', R_{N-2},t |n_1, n_2 \ldots
\rangle=
  \sum_f \sqrt{\frac{n_f}{N}} \: \langle\textbf{r},t | f\rangle \:
  \langle \textbf{r}', R_{N-2},t |n_1, \ldots (n_f-1),\ldots \rangle=$$
$$ =\sum_f \sum_{f', {f'\ne f}}
  \sqrt{\frac{n_f}{N}} \sqrt{\frac{n_{f'}}{N-1}} \:
  \langle\textbf{r},t | f\rangle \: \langle\textbf{r}',t | f'\rangle
  \times$$
$$  \times
  \langle R_{N-2},t |n_1, \ldots (n_{f'}-1),\ldots (n_f-1), \ldots \rangle$$
\begin{equation}  +\sum_f
  \sqrt{\frac{n_f(n_f-1)}{N(N-1)}} \:
  \langle\textbf{r},t | f\rangle \: \langle\textbf{r}',t | f\rangle
  \: \langle R_{N-2},t |n_1, \ldots (n_f-2), \ldots \rangle.
\label{tp cap2:permanent}
\end{equation}
\end{widetext}
Where $\langle \textbf{r},t|f\rangle =\varphi_f(\textbf{r},t)$ ---
one-particle wave function.

 The first term in formula (\ref{tp
cap2:permanent}) represents the particles situating in two
different quantum states, while the second term is referred to
particles in the same quantum state. Therefore, for the particles
in the BEC state, it is sufficient to take into account the second
term in formula (\ref{tp cap2:permanent}). In consideration of the
system of bosons with the temperature differing from zero, where
the certain number of the particles is out of the condensate, the
first summand of formula (\ref{tp cap2:permanent}) gives the
contribution both in the case of interaction of excited particles
with each other and in the case of their interaction with the
particles appearing in the BEC state. In this case, the second
term of formula (\ref{tp cap2:permanent}) gives the contribution
in the interaction both between the particles appearing in the BEC
state and between the excited particles appearing in the same
quantum state.

Moreover using   decomposition formulas  of wave function,
analogously with previous case, for three particle we have:
\begin{widetext}
\begin{equation} \langle \textbf{r}, \textbf{r}',\textbf{r}'', R_{N-3},t |n_1,
n_2 \ldots \rangle=\end{equation}
$$ =\sum_f \sum_{f', {f'\ne f}}\sum_{f''\neq f,f'}
  \sqrt{\frac{n_f}{N}} \sqrt{\frac{n_{f'}}{N-1}} \sqrt{\frac{n_{f''}}{N-2}}\:
  \langle\textbf{r},t | f\rangle \: \langle\textbf{r}',t |
  f'\rangle\: \langle\textbf{r}'',t | f''\rangle
  \times $$
$$ \times
  \langle R_{N-3},t |n_1, \ldots (n_{f''}-1),\ldots (n_{f'}-1),\ldots (n_f-1), \ldots \rangle $$
$$+\sum_f \sum_{f'\neq f}
  \sqrt{\frac{n_f(n_f-1)n_{f'}}{N(N-1)(N-2)}} \:\langle\textbf{r},t | f\rangle \Biggl(\: \langle\textbf{r}',t |
  f'\rangle\: \langle\textbf{r}'',t | f\rangle+\: \langle\textbf{r}',t |
  f\rangle\: \langle\textbf{r}'',t | f'\rangle\Biggr)\times $$
$$ \times\langle R_{N-3},t |n_1, \ldots (n_{f'}-1), \ldots (n_f-2), \ldots
  \rangle $$
$$+\sum_{f} \sum_{f'\neq f}
  \sqrt{\frac{n_f(n_f-1)n_{f'}}{N(N-1)(N-2)}} \:\langle\textbf{r},t | f'\rangle \: \langle\textbf{r}',t |
  f\rangle\: \langle\textbf{r}'',t | f\rangle\times $$
$$\times\langle R_{N-3},t |n_1, \ldots (n_{f}-2), \ldots (n_{f'}-1), \ldots
  \rangle $$
$$  +\sum_f
  \sqrt{\frac{n_f(n_f-1)(n_f-2)}{N(N-1)(N-2)}} \:
  \langle\textbf{r},t | f\rangle \: \langle\textbf{r}',t |
  f\rangle\: \langle\textbf{r}'',t | f\rangle\times $$
\begin{equation} \times\langle R_{N-3},t |n_1, \ldots (n_f-3), \ldots
  \rangle.
  \label{tp cap2:permanent 4}
\end{equation}
\end{widetext}

The first term in formula (\ref{tp cap2:permanent 4}) represents
the particles situating in three different quantum states. The
second and third terms in (\ref{tp cap2:permanent 4}) represents
case when  two particles situating  in one quantum state and one
particle situating in another quantum state. The last term
represents the three particles situating in the same quantum
state. For the particles in the BEC state, it is sufficient to
take into account the last term in (\ref{tp cap2:permanent 4}).
For the consideration of the system of bosons with the temperature
different from zero, where the significant number of the particles
is out of the condensate or in the case of the temperature when
there are no macroscopic number of particles in BEC state, the
terms in formula (\ref{tp cap2:permanent 4}) has the following
influence. The first summand of formula (\ref{tp cap2:permanent
4}) gives the contribution both in the case of interaction of
excited particles with each other and in the case of interaction
of two excited particles with the particles appearing in the BEC
state. The second and the third summands represent case which one
from two states may be BEC state, but another is always excited
state. The last term of formula (\ref{tp cap2:permanent 4}) gives
the contribution in the interaction both between the particles
appearing in the BEC state and between the excited particles
appearing in the same quantum state.

Using relation of orthogonality
\begin{widetext}
$$  \langle n_1,\ldots (n_{f''}-1), \ldots (n_{f'}-1), \ldots
(n_f-1), \ldots |
  n_1, \ldots (n_{q''}-1), \ldots (n_{q'}-1),\ldots (n_q-1),\ldots \rangle=$$
$$  =\delta (f-q)\delta (f'-q') \delta
(f''-q'')+\delta
(f-q)\delta (f'-q'')\delta (f''-q')+ $$

$$+\delta (f-q')\delta (f'-q) \delta
(f''-q'')+\delta (f-q')\delta (f'-q'')\delta (f''-q)+$$
$$+\delta (f-q'')\delta (f'-q) \delta
(f''-q')+\delta (f-q'')\delta (f'-q')\delta (f''-q)$$
and
$$  \langle n_1, \ldots (n_{f'}-1), \ldots (n_f-1), \ldots |
  n_1, \ldots (n_{q'}-1),\ldots (n_q-1),\ldots \rangle=$$
$$\qquad\qquad\qquad  =\delta (f-q) \delta (f'-q')+\delta
(f-q')\delta (f'-q), $$
\end{widetext}
and
$$ \langle n_1, \ldots (n_f-2),\ldots |
  n_1, \ldots (n_q-2),\ldots \rangle=\delta (f-q),
$$
we obtain the expression for two- and three-particle concentration
presented in Sect.3.

\section{\label{sec:level1}Appendix 2}

Details of calculations of $\wp_{n}(\textbf{r},t)$ and
$\hat{m}_{n}(\textbf{r},t)$ are presented here. The quantity
$\wp_{n}(\textbf{r},t)$ and $\hat{m}_{n}(\textbf{r},t)$ reads

$$\wp_{n}(\textbf{r},t)=\sum_{g\neq g_{0}}n_{g}(n_{g}-1)|\varphi_{g}(\textbf{r},t)|^{4}$$
and
$$\hat{m}_{n}(\textbf{r},t)=\sum_{g\neq g_{0}}n_{g}(n_{g}-1)(n_{g}-2)|\varphi_{g}(\textbf{r},t)|^{6}.$$
where $g_{0}$ is a label for state with lowest energy and coincide to the BEC state.

We can calculate $\wp_{n}(\textbf{r},t)$ and $\hat{m}_{n}(\textbf{r},t)$ approximately.

We use method of calculation of correlations described in papers
~\cite{Andreev PRA08} and ~\cite{Maksimov 99}. At the first step
we suppose the particles is free. In this case motion of particles
described by plane-wave.

For the case of plane waves
\begin{equation}\label{tp pl.wave view}\varphi_{p}(\textbf{r},t)=\frac{1}{\sqrt{V}}\exp\biggl(-\frac{\imath}{\hbar}(\varepsilon_{p}t-\textbf{p}\textbf{r})\biggr),\end{equation}
where $\varepsilon_{p}$- is the energy of wave with momentum
$\textbf{p}$. In this case we have
$|\varphi_{g}(\textbf{r},t)|=1/\sqrt{V}$. Consequently, for
$\wp_{n}(\textbf{r},t)$, in the many-particle system we get

$$\wp_{n}(\textbf{r},t)=\frac{1}{V^{2}}\sum_{g}n_{g}(n_{g}-1)$$
$$=\frac{\textit{g}m^{3/2}}{\sqrt{2}\pi^{2}\hbar^{3}V}\int
\sqrt{\varepsilon}d\varepsilon
n_{\varepsilon}(n_{\varepsilon}-1)$$
\begin{equation}\label{tp two-P corr n}\simeq\frac{\textit{g}m^{3/2}}{\sqrt{2}\pi^{2}\hbar^{3}V}\int
\sqrt{\varepsilon}d\varepsilon n_{\varepsilon}^{2},\end{equation}
where $\textit{g}$ is the Lande factor.

The quantum number $g$ indicating quantum state is the energy
$\varepsilon$. Therefore, $n_{\varepsilon}$- is the Bose
distribution function, it's evident form is
\begin{equation}\label{tp }n_{g}=\bar{n}_{g}=\frac{1}{e^{(\varepsilon_{g}-\mu)/T}-1},\end{equation}
where $T$-is the temperature, $\mu$-is the chemical potential,
$$\sum_{g}\bar{n}_{g}=N$$
and
$$\varepsilon_{g}=\frac{p_{g}^{2}}{2m}.$$

Analogously, we write for $\hat{m}_{n}(\textbf{r},t)$
$$\hat{m}_{n}(\textbf{r},t)=\frac{1}{V^{3}}\sum_{g}n_{g}(n_{g}-1)(n_{g}-2)$$
$$=\frac{\textit{g}m^{3/2}}{\sqrt{2}\pi^{2}\hbar^{3}V^{2}}\int
\sqrt{\varepsilon}d\varepsilon
n_{\varepsilon}(n_{\varepsilon}-1)(n_{\varepsilon}-2)$$
\begin{equation}\label{tp TP corr n}\simeq
\frac{\textit{g}m^{3/2}}{\sqrt{2}\pi^{2}\hbar^{3}V^{2}}\int
\sqrt{\varepsilon}d\varepsilon n_{\varepsilon}^{3}.\end{equation}

The quantity $\wp_{n}$ ($\hat{m}_{n}$) expressed in terms of sum
of squares (cubes) of occupation numbers of quantum states
(\ref{tp two-P corr n}), (\ref{tp TP corr n}). Since sum of
squares (cubes) is much less than square (cube) of sum, we obtain
$\wp_{n}\ll n_{n}^{2}$, $\hat{m}_{n}\ll n_{n}^{3}$. Thus, we
neglect by $\wp_{n}$ and $\hat{m}_{n}$ in equation of quantum
hydrodynamics (\ref{tp moment with temp norm}). In formal way, we
can write
$$\begin{array}{ccc}\wp_{n}=0,& \hat{m}_{n}=0.&\end{array}$$

\section{\label{sec:level1}Appendix 3}

Analogously to section (5) we can enter the macroscopic
one-particle wave function for each subsystem of bosons, for BEC
\begin{equation}\label{tp WF in m for B}
\Phi_{B}(\textbf{r},t)=\sqrt{n_{B}(\textbf{r},t)}\exp\biggl(\frac{\imath}{\hbar}m\phi_{B}(\textbf{r},t)\biggr)
\end{equation}
and for noncondensate bosons
\begin{equation}\label{tp WF in m for n}
\Phi_{n}(\textbf{r},t)=\sqrt{n_{n}(\textbf{r},t)}\exp\biggl(\frac{\imath}{\hbar}m\phi_{n}(\textbf{r},t)\biggr)
.\end{equation}

Using equations (\ref{tp cont B}), (\ref{tp cont n}), (\ref{tp moment with temp with corr BEC})
and (\ref{tp moment with temp with corr norm}) we have gotten NLSEs for
each type of bosons. For particles in BEC state the NLSE arise in the form

\begin{widetext}
$$\imath\hbar\partial_{t}\Phi_{B}(\textbf{r},t)=\Biggl(-\frac{\hbar^{2}\nabla^{2}}{2m}+\mu_{B}(\textbf{r},t)+V_{ext}(\textbf{r},t)
-\Upsilon\Biggl(n_{B}+\frac{1}{2}n_{n}+\frac{1}{2}\int\frac{n_{n}}{n_{B}}dn_{B}\Biggr)$$
\begin{equation}\label{tp u GP BEC Tn=0} -\chi\Biggl(n_{B}^{2}+4n_{n}^{2}+4\int(\frac{n_{n}^{2}}{n_{B}}dn_{B}+\frac{2}{3}n_{B}dn_{n}+\frac{4}{3}n_{n}dn_{B})\Biggr)\Biggr)\Phi_{B}(\textbf{r},t)
\end{equation}
and, for noncondensate component NLSE reads
$$\imath\hbar\partial_{t}\Phi_{n}(\textbf{r},t)=\Biggl(-\frac{\hbar^{2}\nabla^{2}}{2m}+\mu_{n}(\textbf{r},t)+V_{ext}(\textbf{r},t)
-\Upsilon\Biggl(\frac{1}{2}n_{B}+2n_{n}+\int\frac{n_{B}}{2n_{n}}dn_{n}\Biggr)$$
\begin{equation}\label{tp u GP N Tn=0} -\chi\Biggl(\frac{2}{3}n_{B}^{2}+8n_{B}n_{n}+6n_{n}^{2}+\frac{2}{3}\int\biggl(\frac{n_{B}+12n_{n}}{n_{n}}\biggr)n_{B}dn_{n}\Biggr)\Biggr)\Phi_{n}(\textbf{r},t)
,\end{equation}
\end{widetext}
where
$$n_{B}=\mid\Phi_{B}(\textbf{r},t)\mid^{2}$$
and
$$n_{n}=\mid\Phi_{n}(\textbf{r},t)\mid^{2}.$$

Wave functions $\Phi_{B}(\textbf{r},t)$ and
$\Phi_{n}(\textbf{r},t)$ are normalized with conditions:
$$ \int d\textbf{r}\Phi_{B}(\textbf{r},t)^{*}\Phi_{B}(\textbf{r},t)=N_{B} ,$$
$$ \int d\textbf{r}\Phi_{n}(\textbf{r},t)^{*}\Phi_{n}(\textbf{r},t)=N_{n} .$$
where $N_{B}$ and $N_{n}$ are number of particles in the BEC
condition and in the noncondensate particles.

\end{document}